\documentclass[12pt]{article}
\usepackage{epsfig}

\def\beq{\begin{equation}}
\def\eeq#1{\label{#1}\end{equation}}
\def\eeqn{\end{equation}}

\def\beqa{\begin{eqnarray}}
\def\eeqa#1{\label{#1}\end{eqnarray}}
\def\eeqan{\end{eqnarray}}

\let\bar=\overbar

\def\Dslash{\not{\hbox{\kern-4pt $D$}}}
\def\dslash{\not{\hbox{\kern-2pt $\del$}}}

\def\msb{{\bar{\ssstyle M \kern -1pt S}}}

\usepackage{fancyhdr,graphicx}
\def\Title#1{\begin{center} {\Large {\bf #1} } \end{center}}

\begin{document}

\Title{Describing SGRs/AXPs as fast and magnetized white dwarfs}

\bigskip\bigskip


\begin{raggedright}

{\it M. Malheiro$^*$ and J. G. Coelho\index{Vader, D.}\\
Instituto Tecnol\'{o}gico de Aeron\'{a}utica\\
Departamento de Ci\^{e}ncia e Tecnologia Aeroespacial\\
12228-900 Vila das Ac\'{a}cias\\
S\~ao Jos\'{e} dos Campos, SP\\
Brazil\\
{\tt $^*$Email: malheiro@ita.br}}
\bigskip\bigskip
\end{raggedright}

\section{Introduction}

Over the last decade, observational evidence has mounted that SGRs/AXPs belong to a particular class of pulsars.
Furthermore, fast and very magnetic white dwarfs have been observed, and 
recently two SGRs with low magnetic fields $B\sim(10^{12}-10^{13})$ G, namely SGR 0418+5729
and Swift J1822.3-1606 were discovered with a period of $P\sim9.08$ s and $P\sim8.44$ s, respectively~\cite{NandaRea,NandaRea2}. 
These new discoveries opens the question of the nature of SGRs/AXPs, emerging alternative scenarios, in particular
the white dwarf (WD) pulsar model~\cite{MMalheiro,Coelho,Coelho2,Coelho3}. These astronomical
observations have based an alternative description of the SGRs/AXPs expressed on rotating highly 
magnetized and very massive WDs (see Malheiro et al. 2012 for more details of this model~\cite{MMalheiro}).

As pointed out in~\cite{Coelho2}, in this new description, several observational
properties are easy understood and well explained as a consequence of the large radius of a massive
white dwarf that manifests a new scale of mass density, moment of inertia, rotational energy, and magnetic dipole moment in
comparison with the case of neutron stars. 

In this contribution, we will show that these recent discoveries of SGRs with low
magnetic field share some properties with the recent detected fast WD pulsar AE
Aquarii, and also with RXJ 0648.0-4418, and EUVE J0317-855, supporting the un-
derstanding of at least these SGRs with low-B as belonging to a class of very fast and
magnetic massive WDs. Furthermore, these recent astronomical observations suggest
that we should revisit the real nature of AXP/SGRs: are they really magnetars or
very fast and massive white dwarfs? In the next section, we present a overview about
this model, and in Section 3 we discuss the recent observations of SGRs with low
magnetic field and the recent observations of fast and magnetized white dwarfs, and
in particular the AE Aquarii as the first white dwarf of a new family of spin-powered
white dwarf pulsars.

\section{A new interpretation - white dwarf pulsar model}

As already discussed in Coelho \& Malheiro 2012~\cite{Coelho,Coelho2}, the magnetic field at the magnetic pole $B_p$ of the star
is related to the dipole magnetic moment by,
\begin{equation}\label{magnetic_moment}
\mid\overrightarrow{m}\mid= \frac{B_p R^3}{2},
\end{equation}
where $R$ is the star radius. If the star magnetic dipole moment is misaligned with the spin axis
by an angle $\alpha$, electromagnetic energy is emitted at
a rate (see e.g., Shapiro and Teukolsky~\cite{shapiro} and references therein),
\begin{equation}\label{edot}
\dot{E}_{\rm dip}= -\frac{2}{3c^3}\mid \ddot{m}\mid^2= -\frac{2\mid\overrightarrow{m}\mid^2}{3c^3}\omega^4\rm sin^2\alpha,
\end{equation}
where $\omega$ is the star angular rotational frequency. Thus, it is the magnetic dipole moment of the star, the physical
quantity that dictates the scale of the electromagnetic radiated power emitted, besides
with the angular rotational frequency. The fundamental physical idea of the rotation-powered
pulsar is that the X-ray luminosity - produced by the dipole field - can be expressed
as originated from the loss of rotational energy of the pulsar,
\begin{equation}\label{erot}
 \dot{E}_{\rm rot}=-4\pi^2I \frac{\dot{P}}{P^3},
\end{equation}
associated to its spin-down rate $\dot{P}$, where $P$ is the rotational period and $I$ is
the momentum of inertia.

Thus, equaling Eqs.~(\ref{edot}) and (\ref{erot}) we deduce the expression of pulsar
magnetic dipole moment,
\begin{equation}\label{moment}
 m=\left(\frac{3c^3I}{8\pi^2}P\dot{P}\right)^{1/2}.
\end{equation}

From Eq.~(\ref{magnetic_moment}) we obtain the magnetic field at the equator $B_e$ as
\cite{Ferrari&Ruffini_1969}
\begin{equation}\label{MagneticField}
 B_e=B_p/2=\left(\frac{3c^3I}{8\pi^2R^6}P\dot{P}\right)^{1/2},
\end{equation}
where $P$ and $\dot{P}$ are observed properties and the moment
of inertia $I$ and the radius $R$ of the object model dependent quantities. 
The description commonly addressed as
magnetar model~\cite{Duncan,Thompson} is based on a canonical neutron star of $M=1.4M_{\odot}$ and $R=10$
km and then $I\sim 10^{45}\rm{g}$ $\rm{cm^2}$ as the source of SGRs and AXPs.
From Eqs.~(\ref{moment}) and (\ref{MagneticField}), and using the parameters above, we obtain the
magnetic dipole moment and the magnetic field
of the neutron star, respectively,
\begin{equation}\label{m_NS}
 m_{\rm NS}=3.2\times10^{37}(P\dot{P})^{1/2} \rm{emu},
\end{equation}
and
\begin{equation}\label{B_NS}
 B_{\rm NS}=3.2\times10^{19}(P\dot{P})^{1/2} \rm{G}.
\end{equation}

For the case of the white dwarf model we use a
radius $R= 3000$ km for all SGRs and AXPs and a mass $M=1.4M_{\odot}$, as recent studies
of fast and
very massive white dwarfs obtained (see K. Boshkayev et al. 2012~\cite{JRueda}). Thus, these values
of mass and radius generating the momentum of inertia $I\sim 1.26\times 10^{50}\rm{g}$ $\rm{cm^2}$,
will be adopt hereafter in this work as the fiducial white dwarf model parameters.
Using that parameters we obtain the
magnetic dipole moment and the magnetic field
of the white dwarf pulsar, respectively,
\begin{equation}\label{m_WD}
 m_{\rm WD}=1.14\times10^{40}(P\dot{P})^{1/2} \rm{emu},
\end{equation}
and
\begin{equation}\label{B_WD}
 B_{\rm WD}=4.21\times10^{14}(P\dot{P})^{1/2} \rm{G}.
\end{equation}

These results clearly shows that the scale of the dipole magnetic moment in WD is $\sim10^3$ times larger than for neutron stars, 
exactly the factor seen in the X-ray luminosity of SGRs/AXPs when compared with $L_X$ of slow pulsars ($P\sim$1 to 10 s)
as X-ray Dim isolated neutron stars (XDINs) and high-B radio pulsars (see Coelho \& Malheiro 2012~\cite{Coelho2}).
Furthermore, the surface
magnetic field of WDs is $\sim10^5$ smaller than the ones of neutron stars, eliminating all the overcritical $B$ fields deduced in 
the magnetar model. Then, the basic idea is that, being a WD $\sim10^3$ times bigger than a NS, at comparable mass, its
moment of inertia is $\sim10^5$ times larger. This implies that the rotational energy lost can be large enough
to explain the observed X-ray luminosity in SGRs/AXPs ($\sim10^{32}-10^{36}$ erg/s) even for quite low values
of the period derivative $\dot{P}$.

\section{Observations of magnetized white dwarfs and SGRs with low B}

Magnetized white dwarfs (MWDs) constitute at least 10\% of the white dwarfs if observational biases are
considered~\cite{Kawka}. The current known population of MWDs has been increased considerably 
by the Sloan Digital Sky Survey (SDSS) to about 220 objects (see K$\rm \ddot{u}$lebi et al. 2013~\cite{Kulebi2013} for more details). 
SDSS also dramatically increased the total known white dwarf 
population (see e.g. Kleinman et al. 2013, for Data Release 7~\cite{Kepler}) and recent studies indicate that the number of MWDs 
in the SDSS could be as large as 521~\cite{Kepler2,Kulebi2013}.
Furthermore, some sources have even been tentatively proposed as candidates for white dwarf pulsars. A specific example is 
AE Aquarii, the first white dwarf pulsar, very fast with a short period $P= 33.08$ s~\cite{Terada2,Angel}. 
The rapid braking of the white dwarf and the nature of pulse hard X-ray emission
detected with japanese SUZAKU space telescope under these conditions can be explained in
terms of spin-powered pulsar mechanism.
Although AE Aquarii is in a binary system with orbital period $\sim9.88$ hr, and not an isolated pulsar: very likely the power due to accretion of matter is
inhibited by the fast rotation of the white dwarf~\cite{Ikhasanov4}.

Recently, Mereghetti et al. 2009~\cite{Mereghetti3}
showed that the X-ray pulsator RX J0648.0-4418 is a white dwarf with mass $M=1.28M_\odot$ and radius
$R= 3000$ km, and spin period $P = 13.2$ s. 
EUVE J0317-855, is another WD pulsar candidate discovered recently~\cite{Kulebi}. However, relevant pulse emission 
has not been observed yet, which may suggest that the electron-positron creation
and acceleration does not occur (see Kashiyama et al.~\cite{Kashiyama}). Barstow et al. 1995~\cite{Barstow} obtained a period
of $P\sim$ 725 s, which is also a fast and very magnetic WD with a dipole magnetic field is
$B\sim 4.5\times 10^8$ G (obtained by optical photometric and polarimetric), 
and a mass $(1.31 - 1.37)M_\odot$ which is relatively large
compared with the typical WD mass $\sim 0.6 M_\odot$. In this work, we describe AE Aquarii and RX J0648.0-4418 as
a rotation powered white dwarf, and obtain the magnetic field,
using the  white dwarf parameters presented in the last section.

\begin{figure}[!]
\begin{center}
\epsfig{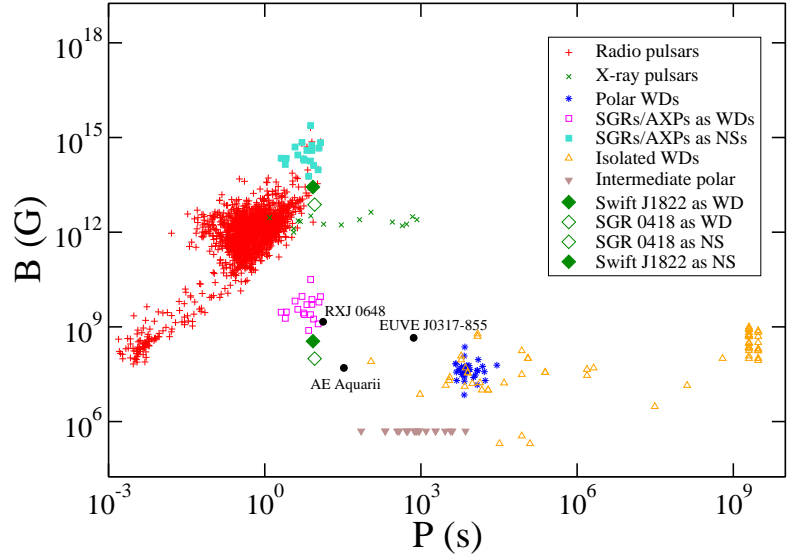}
\caption{The figure shows the magnetic field strength 
of neutron stars and magnetic white dwarfs as a function of the
rotational period.}
\label{fig1}
\end{center}
\end{figure}

In Fig.~\ref{fig1}, we present the magnetic field as a function of the period of these two SGRs with low B described 
in the white dwarf model with the three fast white dwarfs presented above. 
Here we see that the magnetic field $B$ of the SGRs and AXPs as NSs or WDs are quite different ($\sim 10^5$ order
of difference) as explained in the last section, and already pointed out in 
 Refs.~\cite{MMalheiro, Coelho, Coelho2}. The magnetic field $B$ of the two
recent SGRs described as white dwarf pulsars are comparable with the ones observed
for the fast white dwarfs also plotted. The magnetic WDs shown in Fig.~\ref{fig1}, are the
complete sample obtained by SDSS project, for which the period $P$ and magnetic field
$B$ are known. The magnetic fast white dwarfs are separated in two classes: isolated and polars, very magnetic
with $B\sim (10^7-10^8)$ G, and the intermediate polars with weaker field $B\sim 10^5$ G.

In Table~\ref{ta1} we compare and contrast the parameters of these three fast white dwarfs with the two SGRs with low B described 
in the white dwarf model presented before. 
As already shown in Ref.~\cite{Coelho, Coelho2}, important features of the two SGRs with low magnetic field are
very similar to the ones of fast and magnetic white dwarfs recently detected. They are old,
characteristic ages of Myr, low quiescent X-ray luminosity $L_X \sim (10^{30}-10^{32})$,
magnetic field of $B_{\rm WD}\sim (10^7-10^8)$ G and magnetic dipole moments of $m_{\rm WD}\sim (10^{33}-10^{34})$ emu.
These results give evidence for the interpretation of SGRs/AXPs as being rotating white dwarf pulsars, at least the two
SGRs with low-B.

\begin{table}[htb!]
\begin{center}
\setlength{\tabcolsep}{2pt} 
\begin{tabular}{l|c|l|c|l|c}\hline
& SGR 0418 &Swift J1822 &AE Aquarii &RXJ 0648 &EUVE J0317 \\ \hline
$P$(s) &9.08 &8.44 & 33.08 &13.2 &725\\
$\dot{P}$ $(10^{-14})$ &0.4 &8.3 &5.64 &$<$ 90 &- \\
$\rm Age$ $(\rm Myr)$ &24 &1.6 &9.3 &0.23 &- \\
$L_X$ $(\rm erg/s)$ &$\sim1.0\times10^{30}$ &$\sim4.2\times10^{32}$ &$\sim10^{31}$ &$\sim10^{32}$ &- \\
$B_{\rm WD}$(G) &$\sim8.02\times10^7$ &$\sim3.52\times10^{8}$ &$\sim 5\times10^7$ &$<$ $1.45\times10^9$ &$\sim 4.5\times10^8$ \\
$B_{\rm NS}$(G) &$\sim6.10\times10^{12}$ & $\sim2.70\times10^{13}$ &- &- &- \\
$m_{\rm WD}$(emu) &$\sim2.17\times10^{33}$ &$\sim0.95\times10^{34}$ &$\sim1.35\times10^{33}$ &$3.48\times10^{34}$ &$1.22\times10^{34}$ \\
$m_{\rm NS}$(emu) &$\sim6.10\times10^{30}$ &$\sim2.70\times10^{31}$ &- &- &- \\ \hline
\end{tabular}
\caption{Comparison of the observational properties of five sources: SGR 0418+5729 and Swift J1822.3-1606 (see N. Rea et al. 2010, 2012)
and three observed white dwarf pulsar candidates. For the SGR 0418+5729 and Swift J1822.3-1606 the parameters
$P$, $\dot{P}$ and $L_X$ have been taken from the McGill
online catalog at www.physics.mcgill.ca/~pulsar/magnetar/main.html. The characteristic age is given
by Age $=P/2\dot{P}$ and the magnetic moment $m$ and
the surface magnetic field $B$ are given by Eqs.~(\ref{moment}) and (\ref{MagneticField}), respectively.}
\label{ta1}
\end{center}
\end{table}
\newpage
JGC and MM acknowledges the Brazilian agency FAPESP (thematic project 2007/03633-3), CNPq and CAPES. We are grateful to Y. Terada by the data points of
the figure in this paper, and also by the valuable discussions.

\end{document}